\documentclass[amsmath,amssymb,aps,pra,superscriptaddress,twocolumn,notitlepage]{revtex4-1}
\usepackage{natbib}
\usepackage{ color}
\usepackage{amssymb}
\usepackage{amsmath, amsthm}
\usepackage{epsfig}
\usepackage{epstopdf}
\usepackage{graphicx}
\usepackage{bbm}
\usepackage{changes}

\newcommand{\I}{\mathbb I}

\newcommand{\ket}[1]{\vert#1\rangle}
\newcommand{\bra}[1]{\langle#1\vert}

\usepackage{graphicx}
\usepackage{bm}

\newcommand{\ba}{\begin{eqnarray}}
\newcommand{\be}{\begin{equation}}
\newcommand{\ee}{\end{equation}}
\newcommand{\beq}{\begin{equation}}
\newcommand{\eeq}{  \end{equation}}
\newcommand{\bea}{\begin{eqnarray}}
\newcommand{\eea}{  \end{eqnarray}}
\newcommand{\ea}{\end{eqnarray}}
\newcommand{\ban}{\begin{eqnarray*}}
\newcommand{\ean}{\end{eqnarray*}}

\newcommand{\ie}{{\it{i.e.}~}}

\begin{document}

\title{Experimental investigation of partially entangled states for device-independent randomness generation and self-testing protocols}

\begin{abstract}

Previous theoretical works showed that all pure two-qubit entangled states can generate one bit of local randomness and can be self-tested through the violation of proper Bell inequalities. We report an experiment in which nearly pure partially entangled states of photonic qubits are produced to investigate these tasks in a practical scenario. We show that small deviations from the ideal situation make low entangled states impractical to self-testing and randomness generation using the available techniques. Our results show that in practice lower entanglement implies lower randomness generation, recovering the intuition that maximally entangled states are better candidates for device-independent quantum information processing.
\end{abstract}


\author{S. G\'omez}
\affiliation{Departamento de F\'isica, Universidad de Concepci\'on, 160-C Concepci\'on, Chile}
\affiliation{Millennium Institute for Research in Optics, Universidad de Concepci\'on, 160-C Concepci\'on, Chile.}
\author{A. Mattar}
\affiliation{ICFO-Institut de Ciencies Fotoniques, The Barcelona Institute of Science and Technology, 08860 Castelldefels, Barcelona, Spain}
\author{I. Machuca}
\affiliation{Departamento de F\'isica, Universidad de Concepci\'on, 160-C Concepci\'on, Chile}
\affiliation{Millennium Institute for Research in Optics, Universidad de Concepci\'on, 160-C Concepci\'on, Chile.}
\author{{E. S. G\'omez}}
\affiliation{Departamento de F\'isica, Universidad de Concepci\'on, 160-C Concepci\'on, Chile}
\affiliation{Millennium Institute for Research in Optics, Universidad de Concepci\'on, 160-C Concepci\'on, Chile.}
\author{D. Cavalcanti}
\affiliation{ICFO-Institut de Ciencies Fotoniques, The Barcelona Institute of Science and Technology, 08860 Castelldefels, Barcelona, Spain}
\author{O. Jim\'enez Far\'ias}
\affiliation{ICFO-Institut de Ciencies Fotoniques, The Barcelona Institute of Science and Technology, 08860 Castelldefels, Barcelona, Spain}
\author{A. Ac\'in}
\affiliation{ICFO-Institut de Ciencies Fotoniques, The Barcelona Institute of Science and Technology, 08860 Castelldefels, Barcelona, Spain}
\affiliation{ICREA-Instituci\'o Catalana de Recerca i Estudis Avan\c cats, Lluis Companys 23,
08010 Barcelona, Spain}
\author{G. Lima}
\affiliation{Departamento de F\'isica, Universidad de Concepci\'on, 160-C Concepci\'on, Chile}
\affiliation{Millennium Institute for Research in Optics, Universidad de Concepci\'on, 160-C Concepci\'on, Chile.}


\date{\today}


\maketitle

\section{Introduction}
Quantum correlations, a manifestation of entanglement, are a key resource for the implementation of new quantum technologies. In particular, correlations violating Bell inequalities~\cite{Bell_original} are the basis for device-independent (DI) quantum information protocols \cite{Brunner:2014ix}, in which the security and performance of the tasks are inferred solely from the measurement statistics -i.e., with no need for additional assumptions about the working mechanism of the devices. The implementation of DI protocols is challenging, but the recent demonstration of loophole-free Bell tests \cite{Giustina15, Shalm15, hensen15,Rosenfeld17} open new avenues where the scope of DI technologies can be tested.

Two promising applications of DI quantum information protocols are randomness certification and self-testing. The first refers to the fact that a violation of a Bell inequality certifies that the measurement outcomes can be unpredictable and private \cite{PhDColbeck, Pironio:2010kx, acin16}. To date, several implementations on randomness certification have been reported \cite{Pironio:2010kx,Kwiat13,Bierhorst18,Pan18,Pan18Nature,gomez2018}. The second refers to the fact that the maximal violation of certain Bell inequalities can only be obtained by performing specific measurements on specific quantum states  \cite{MY, YN, SCAA, McKYS, CGS, Kaniewski}. Hence, such violation can be used as a DI estimation method.

Both maximal randomness certification and self-testing were proven to be possible to achieve with partially entangled states \cite{Acin12,YN,CGS}. However, such tasks require idealized noiseless situations. This article aims at investigating these protocols in practical situations. We use a high-quality spontaneous parametric-down conversion based source of partially entangled states (PES) to implement both protocols. The observed visibilities are of the order of $99.7\%$, resulting in nearly maximal Bell inequality violations. However, we also observe that the presence of  noise, even if low, drastically reduces the protocol performance for non-maximal entanglement. We therefore see that, in practice, and perhaps not surprisingly, lower entanglement implies lower certified randomness and self-testing fidelities.

\section{Device-independent randomness certification and self-testing}
In the device-independent scenario, no assumptions are made on the states prepared or on the implemented measurement. Instead, measurement apparatuses are treated as black boxes receiving classical inputs (corresponding to the measurement choices) and providing classical outcomes (the measurement results). Here we work on a bipartite scenario involving two of such black-boxes $A$ and $B$. The measurement choices of these boxes are labeled by $x$ and $y$, respectively, and is considered to assume values $0$ or $1$. Similarly, measurement outcomes are labeled $a$ and $b$, respectively, and assume binary values $a,b=\pm 1$. At each round of the experiment, a particular value of $x$ and $y$ is selected and values for $a$ and $b$ are obtained. After many rounds of the experiment, we can estimate the probability distributions $P(ab|xy)$, which are often named simply by correlations.
The correlations $P(ab|xy)$ are said to be non-local if they violate a Bell inequality \cite{Brunner:2014ix}.

\subsection{Randomness certification}
If the correlations $P(ab|xy)$ are nonlocal, the outcomes $a$ (or $b$) can not be predicted \cite{acin16}. This unpredictability can be estimated by assuming the worst-case scenario where the two quantum particles in devices $A$ and $B$ are correlated with a third quantum particle held by an eavesdropper Eve, the global tripartite being pure and denoted by $\ket{\Psi}$. Eve's goal is to guess the output $a$ for a particular measurement choice $x^*$ by performing a measurement on her part of the state. Eve is assumed to know $\ket{\Psi}$ and the measurements implemented by boxes $A$ and $B$. The randomness of $a$ when $x=x^*$ can be estimated through the guessing probability \cite{NPS14,BSS14}:
\begin{align}
\label{eq:randomnessSDP}
&P_{\text{guess}}=\max_{\{\ket{\Psi},\Pi_{a|x},\Pi_{b|y},\Pi_{e}\}} \sum_a\bra{\Psi} \Pi_{a|x^*}\otimes\I\otimes\Pi_{e=a}\ket{\Psi},\\
&\textrm{such that             } P(ab|xy)=\bra{\Psi} \Pi_{a|x}\otimes\Pi_{b|y}\otimes\I\ket{\Psi}\nonumber.
\end{align}
This quantity gives the maximum probability that Eve's outcome $e$ matches the user's outcome $a$ for measurement $x^*$ over all possible quantum realizations, described by a tripartite quantum state $\ket{\Psi}$ and measurements $\Pi_{a|x}$, $\Pi_{b|y}$ and $\Pi_e$ for devices $A$, $B$ and $E$, compatible with the observed distributions $P(ab|xy)$. The estimated randomness can be  expressed in bits by $R=-\log_2 (P_{\text{guess}})$.

The optimisation problem \eqref{eq:randomnessSDP} is typically difficult as variables could be states and measurements of any dimension. However, upper bounds to the solutions can be efficiently computed by the semi-definite programming (SDP) techniques proposed in Refs. \cite{NPS14,BSS14}.

\subsection{Tilted Bell Inequality and Self-testing}

The tilted Bell inequality was introduced in~\cite{Acin12} to study the properties of pure two-qubit entangled states for randomness certification. It reads
\begin{eqnarray}\label{tiltedCHSH}
B_\alpha=\alpha \langle A_0\rangle+ \sum_{j,k=0}^1 (-1)^{j k}\langle A_j\times B_k\rangle \leq \alpha+2,
\end{eqnarray}
where $\langle A_x\rangle=P(+1|x)-P(-1|x)$ and $\langle A_j\times B_k\rangle=P(a=b|xy)-P(a\neq b|xy)$ and $0\leq\alpha\leq 2$. Its maximal quantum violation is equal to $B^{max}_\alpha= \sqrt{8+2\alpha^2}$ and can be obtained by performing appropriate measurements on the states
\begin{eqnarray}
\ket{\psi(\theta)}=\rm{cos}(\theta)\ket{00}+\rm{sin}(\theta)\ket{11}, \label{PES}
\end{eqnarray}
with $\theta=\frac{1}{2}\arctan (\sqrt{\frac{4-\alpha^2}{2\alpha^2}})$. The usefulness of this inequality for randomness certification comes from the fact that, at the maximal quantum violation, the guessing probability for the output of measurement $x=1$ is equal to $1/2$ for all non-zero values of $\alpha$, that is, for any non-zero entanglement in the pure state~\eqref{PES}.

The maximal quantum violation of the inequality \eqref{tiltedCHSH} also implies that the underlying measured state is equivalent, up to local isometries, to the state \eqref{PES} \cite{YN}. To be more specific, upon observation of the value $B^{max}_\alpha$ we can conclude that the state $\rho$ shared by boxes $A$ and $B$ is related to the state \eqref{PES} through local channels,
\begin{eqnarray} \label{statement}
\begin{array}{l}
\Lambda_A\otimes\Lambda_B(\rho) = \ket{\psi(\theta)}\bra{\psi(\theta)},
\end{array}
\end{eqnarray}
where $\Lambda_A$ and $\Lambda_B$ are completely positive maps. In other words, the maximal violation of the inequality implies that we can deterministically and locally transform the actual state shared by the boxes into the desired PES.

In case a non-maximal violation of the inequality \eqref{tiltedCHSH} is observed, a a bound on the minimal overlap between the actual state $\rho$ shared by the boxes and the state \eqref{PES} as a function of the observed Bell violation was obtained in \cite{coopmans17}. It was show that, for angles $\theta \in [0.14, \pi/4]$, the fidelity between the physical state $\rho$ and the state to be self tested (\ref{PES}) can be lower bounded by a linear function of $B_{\alpha}$
\begin{eqnarray} \label{fidelitybound}
\max_{\Lambda_A,\Lambda_B} F((\Lambda_A\otimes\Lambda_B)\rho,\ket{\psi(\theta)}\bra{\psi(\theta)})\geq s_{\alpha}\cdot B_{\alpha} +\mu_{\alpha},
\end{eqnarray}
where $\alpha=2/\sqrt{1+2\rm{tan}^2(2\theta)}$. The parameters $s_{\alpha}$ and $\mu_{\alpha}$ are given by
\begin{eqnarray}
\begin{array}{l}
s_{\alpha}=\frac{1-\frac{1}{4}(1+\sqrt{\frac{4-\alpha^2}{8+2\alpha^2}}+\sqrt{\frac{2\alpha^2}{8+2\alpha^2}} )}{\sqrt{8+2\alpha^{2}}-(2+\alpha)},\\
\mu_{\alpha}= 1-s_{\alpha}\sqrt{8+2\alpha^{2}}.
\end{array}
\end{eqnarray}

In our experiment we use \eqref{fidelitybound} to lower bound the fidelity of the experimentally produced state with a partially entangled state. To do this, we first measure the expectation values entering in (\ref{tiltedCHSH}), and then we optimise the right-hand-side (RHS) of \eqref{fidelitybound} as a function of $\alpha$,
This provides the largest value of the fidelity with a pure entangled two-qubit state that we can guarantee from the observed statistics.

We notice that Ref. \cite{SelfTestingGuo18} reported on a self-testing experiment with partially entangled states. However, this test was based on the Clauser-Horn-Shimony-Holt Bell inequality \cite{CHSH69}, which is not optimal for partially entangled states.

\section{Experiment}

\begin{figure}[t]
\centering
\includegraphics[width=0.5\textwidth]{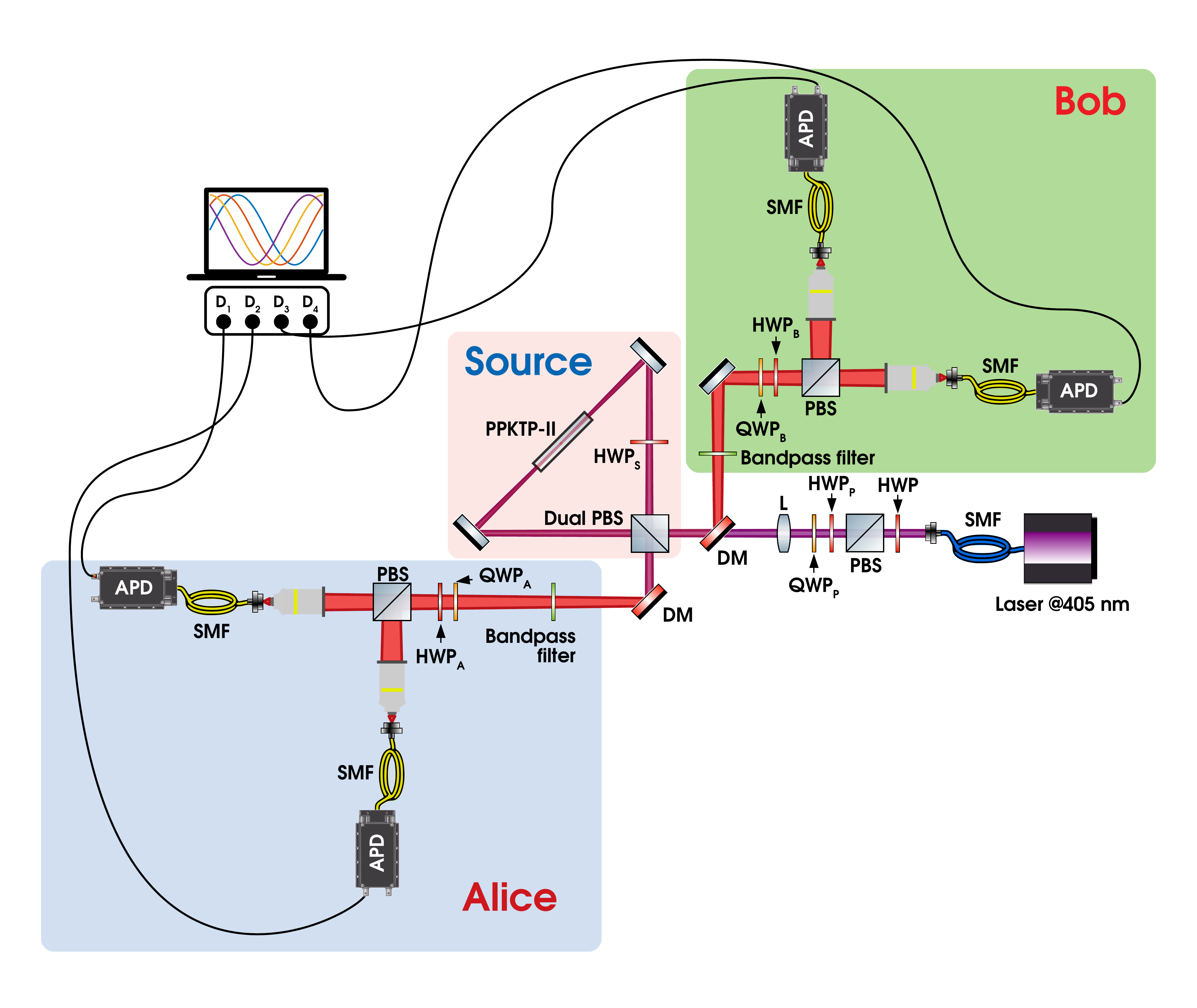}
\caption{Experimental setup used for randomness certification and self-testing. See more details in the main text.} \label{fig:setup}
\end{figure}

To experimentally study randomness certification and self-testing in a Bell scenario with partially entangled states, we employed a high-purity and tunable polarization entanglement source. The experimental setup is depicted in Fig. (\ref{fig:setup}). It consists of an ultrabright source of polarization-entangled photons based on an intrinsically phase-stable Sagnac interferometer and the spontaneous parametric down-conversion (SPDC) process \cite{kim2006a,Kim2006b,Zeilinger2007,gomez2016device}. The observed spectral brightness was ~410000 photon pairs (s mW nm)$^{-1}$. A pigtailed, single mode continuous-wave laser operating at 405 nm is used to excite a type-II nonlinear periodically poled potassium titanyl phosphate (PPKTP-II) 20 mm long crystal generating pairs of polarization-entangled photons at 810 nm. The crystal is placed inside of the interferometer, which is composed of two laser mirrors, a half-wave plate (HWP$_\textrm{S}$), and a polarizing beam-splitter cube (PBS$_\textrm{1}$). The HWP$_\textrm{S}$ and the PBS$_\textrm{1}$ are coated for both pump and down-converted wavelengths. Moreover, the HWP$_\textrm{S}$ is set with its fast axis at $45^\circ$ with respect to the horizontal. In this case, the down-converted photons are generated in the clockwise and counterclockwise directions inside of the interferometer. Then, the clockwise and counterclockwise propagating modes of the generated photons overlap inside the PBS$_\textrm{1}$, resulting in the polarized-entangled state $|\Psi\rangle=\alpha|HV\rangle+\beta|VH\rangle$. The coefficients $\alpha$ and $\beta$ arise from the linear polarization mode of the pump beam $\beta|H\rangle+\alpha|V\rangle$, and therefore the amount of entanglement of the generated state can be adjusted with simple polarization optics at the pump beam propagation path. In our case, partially polarization-entangled states (PES) are generated by properly adjusting the fast axis angles of the HWP$_\textrm{P}$, and a {quarter-wave} plate QWP$_\textrm{P}$. In our study we consider the generation of five PESs with the concurrence equally separated and covering all the concurrence range, as for instance also done in previous experiments~\cite{walborn06b,salles08,lima09b}.

To generate high-quality PES with high fidelities we have used Semrock high-quality narrow bandpass filters [full width at half maximum (FWHM) of 0.5 nm, with a peak transmission $>$ 90\%] centered at 810 nm to guarantee the degenerate generation of the down-converted photons. Also, to avoid distinguishability between the spatial modes of “HV” and “VH”, we resort to a numerical model to optimally coupling the generated polarization-entangled photons into single-mode optical fibers after being transmitted by the PBS$_\textrm{1}$ \cite{Ljunggren_PRA_fibers}. Considering that the beam waist of the pump mode is $w_p$, and that $w_{SPDC}$ is the waist of the down-converted photon modes at the center of the PPKTP-II crystal, the optimal condition for the maximal coupling is obtained when $w_{SPDC}=\sqrt{2}w_p$. In our configuration, we achieve such condition by using a 20 cm focal length lens $L$ to focus the pump beam at the center of the PPKTP-II crystal, and by using a 10$\times$ objective lenses for coupling the down-converted photons into the optical fibers.

To ensure a high two-photon visibility, and therefore a high violation of the inequality (\ref{tiltedCHSH}) for each generated state, the required local projective measurements were performed by Alice (Bob) using high-quality polarizing optics components, composed of a HWP$_\textrm{A}$ (HWP$_\textrm{B}$), a QWP$_\textrm{A}$ (QWP$_\textrm{B}$) and a PBS. Also, we used extra polarizer films at the front of the detectors (not shown in the figure) to guarantee an extinction ratio equal to $10^7:1$. In this way, the contrast of the polarizing optics does not limit the two-photon visibility below 99$\%$. PerkinElmer single-photon avalanche detectors (APDs) with a detection efficiency of 50\% (@ 810nm) were used, resulting on an overall detection efficiency of 15\% (specifically:  42\% of fiber collection efficiency and 72\% of transmission over the filtering and polarizing systems). To reduce the accidental coincidence rate probability, we resort on a high-resolution coincidence electronics to implement 500 ps coincidence window \cite{gomez2018,gomez2016device}. The obtained overall two-photon visibility is (99.7$\pm$0.3)\% while measuring over both the logical and diagonal polarization bases. Note that this high value for the visibility is obtained through post-selection. In fact, the observed Bell violation does not close the detection loophole and is only valid under the fair sampling assumption. Yet, this is enough for our purposes, as our goal is not to implement a fully DI protocol but to study the effect of imperfections on implementations of randomness certification and self-testing protocol using PES.

Finally, our setup has been specially tailored to avoid the apparent signalling effects that arise when there are small drifts in the experiment due to pump power fluctuations. Since SPDC is a non-linear process, the effect of laser fluctuations can be non-negligible on the rate of generating down-converted photons over time, thus mimicking signalling. Each measurement block in our setup lasts 10 seconds. For each setting, ten blocks have been used. To avoid fluctuations in the marginal counts during the data acquisition procedure, we adopted a precise active pump power control of 1 $\mu$W, while the pump beam operates with 1 mW. Thus, with a precision of $10^{-3}$ such control maintains the pump beam power stable for the entire day, which is much longer than the full experiment duration.

\section{Results}

\begin{figure}
\centering
\includegraphics[width = 1\linewidth]{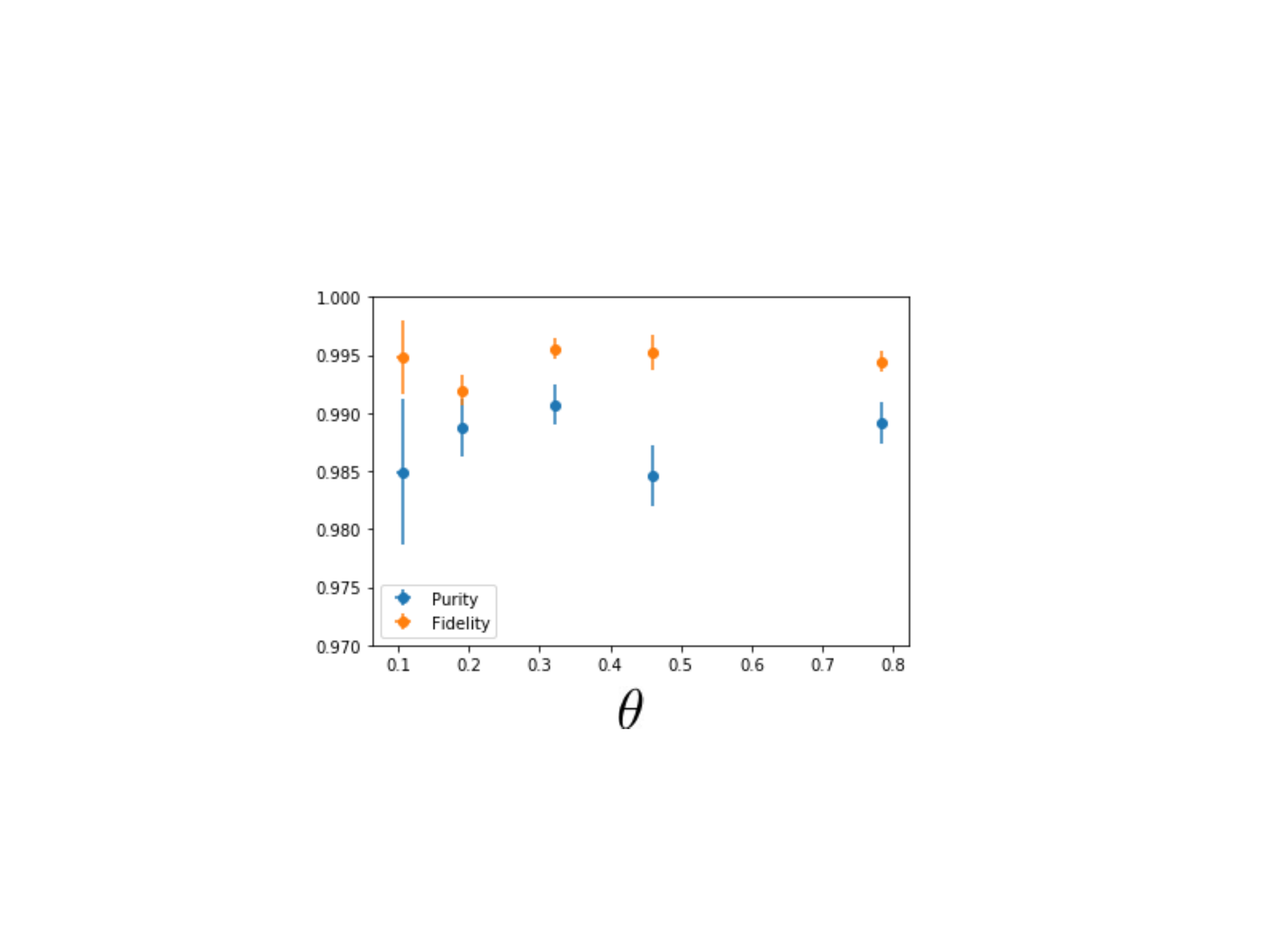}
\caption{Purity Tr$(\rho_t^2)$ (blue points) and the fidelity (orange points) with respect to the closest PES obtained from quantum state tomography. In the horizontal axis the angle $\theta$ characterising  the closest PES. The error bars are obtained with Gaussian error propagation and considering the Poisson statistics of the recorded coincidence counts.}
\label{PurFid}
\end{figure}

After presenting our setup we are now in position to discuss the observed results for the two considered protocols, namely randomness certification and self-testing. However, before doing that,
and for the sake of reference, we first make quantum state tomography of the five states considered in our experiment, assuming our system consists of two qubits. All the tomographic state $\rho_t$ reconstructed present degrees of purity Tr[$\rho^2_t$] above $0.985$ (see Fig. \ref{PurFid}). We also calculate the degree of entanglement of $\rho_t$ as measured by the concurrence. We then proceed to find the closest pure PES, $\ket{\psi(\theta)}$, by maximizing the fidelity
\begin{eqnarray}
F(\rho_t, \ket{\psi(\theta)}\bra{\psi(\theta)})=\bra{\psi(\theta)}\rho_t \ket{\psi(\theta)}.
\end{eqnarray}
This allows to estimate the fidelity with the target PES and the corresponding angle $\theta$. The results are also shown in Fig \ref{PurFid}. As one can see, all reconstructed states have remarkably high fidelities with pure PESs of the form of $\ket{\psi(\theta)}$, all of them being larger than 0.99.

\begin{figure}
\centering
\includegraphics[width = 1\linewidth]{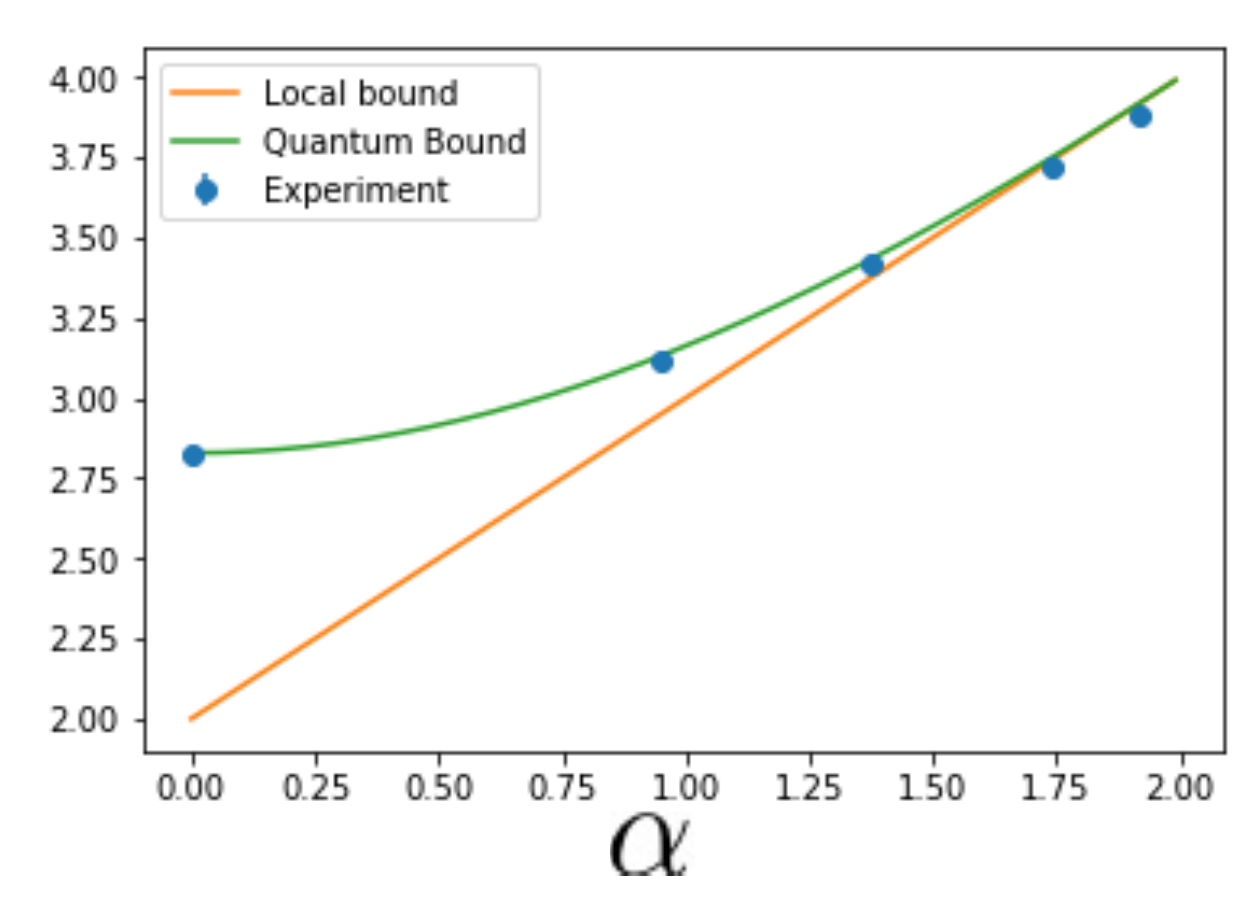}
\caption{Observed violation of the tilted Bell inequality. The orange line corresponds to the local bound, above which nonlocality can be proven.  The green line is the maximal violation that can be achieved, while the blue dots are the experimental values obtained. The error bars lie within the experimental dots and are obtained with Gaussian error propagation and considering the Poisson statistics of the recorded coincidence counts.}
\label{loquant}
\end{figure}

Since this angle $\theta$ characterizes to a good extent the prepared state, we use it to choose the settings in our Bell experiment and set $\alpha=2/\sqrt{1+2\tan^2(2\theta)}$ for the tilted Bell inequality. In our experiment, we measure all the probabilities P$(ab|xy)$, not only those probabilities involved in the inequality, as they will be later used for the randomness analysis. In figure (\ref{loquant}), we plot the obtained values of the Bell parameter $B_{\alpha}$ determined by the tilted Bell inequality. There, it becomes clear that while the observed violation is always very close to the maximal one, the gap with the classical bound closes when increasing (decreasing) the value of $\alpha$ ($\theta$), or equivalently, decreasing the entanglement of the states~\eqref{PES}.

\subsection{Device-independent randomness certification}

With the settings of the experiment tuned to observe the best violation of the inequality \eqref{tiltedCHSH}  we proceed to estimate the probabilities $P(ab|xy)$. Due to finite statistics effects, the obtained set of probabilities present a small amount of signalling. For this reason we use the  Collins-Gisin regularization method to obtain a set of no-signaling probability distributions $P_{NS}(ab|xy)$ that approximate the observed distribution \cite{CollinsGisin04} . With $P_{NS}$ we run SDP optimization introduced in Refs. [10, 11] that provides an upper bound to the guessing probability (1). The solution of this SDP optimization provides a linear function $S(P(ab|xy))$, or simply $S$, whose value is a lower bound on the amount of randomness of any correlations $P(ab|xy)$. We finally rewrite $S$ in terms of expectation values and use it to estimate the amount of randomness in our experiment from the actual measured expectation values. See \cite{Bourdoncle18} for more details of this method. The errors of the recorded probabilities are calculated assuming fair samples from Poissonian distributions and Gaussian error propagation.

\begin{figure}
\centering
\includegraphics[width = 1\linewidth]{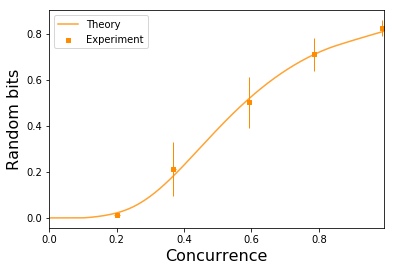}
\caption{Experimental randomness certification as a function of entanglement. Almost all the cases can certify randomness. However, they are far from, in principle, the possible value of 1 bit of randomness. The solid line is given as a reference, and considers PES with $0.5\%$ of white noise.The error bars are obtained with Gaussian error propagation and considering the Poisson statistics of the recorded coincidence counts.}
\label{rand}
\end{figure}

We compare the obtained guessing probabilities with the concurrence computed through tomography. Figure (\ref{rand}) shows the results of our calculations. We observe that randomness decreases drastically as the state is less entangled due to small imperfections of the statistics. Remember that in the absence of noise, the amount of randomness should be equal to 1 bit for all values of entanglement. Note also that even for those cases where the entanglement in the state is very close to one, we are still significantly far from the ideal 1 bit of randomness. This is because the guessing probability is very sensitive to imperfections in the almost noiseless case.

We notice that the function $S(P(ab|xy))$ obtained numerically to lower bound the generated randomness has coefficients that increase when decreasing the entanglement of the state. This explains why the error bars increase in this regime.

\subsection{Self-testing}

\begin{figure}
\centering
\includegraphics[width = 1.0\linewidth]{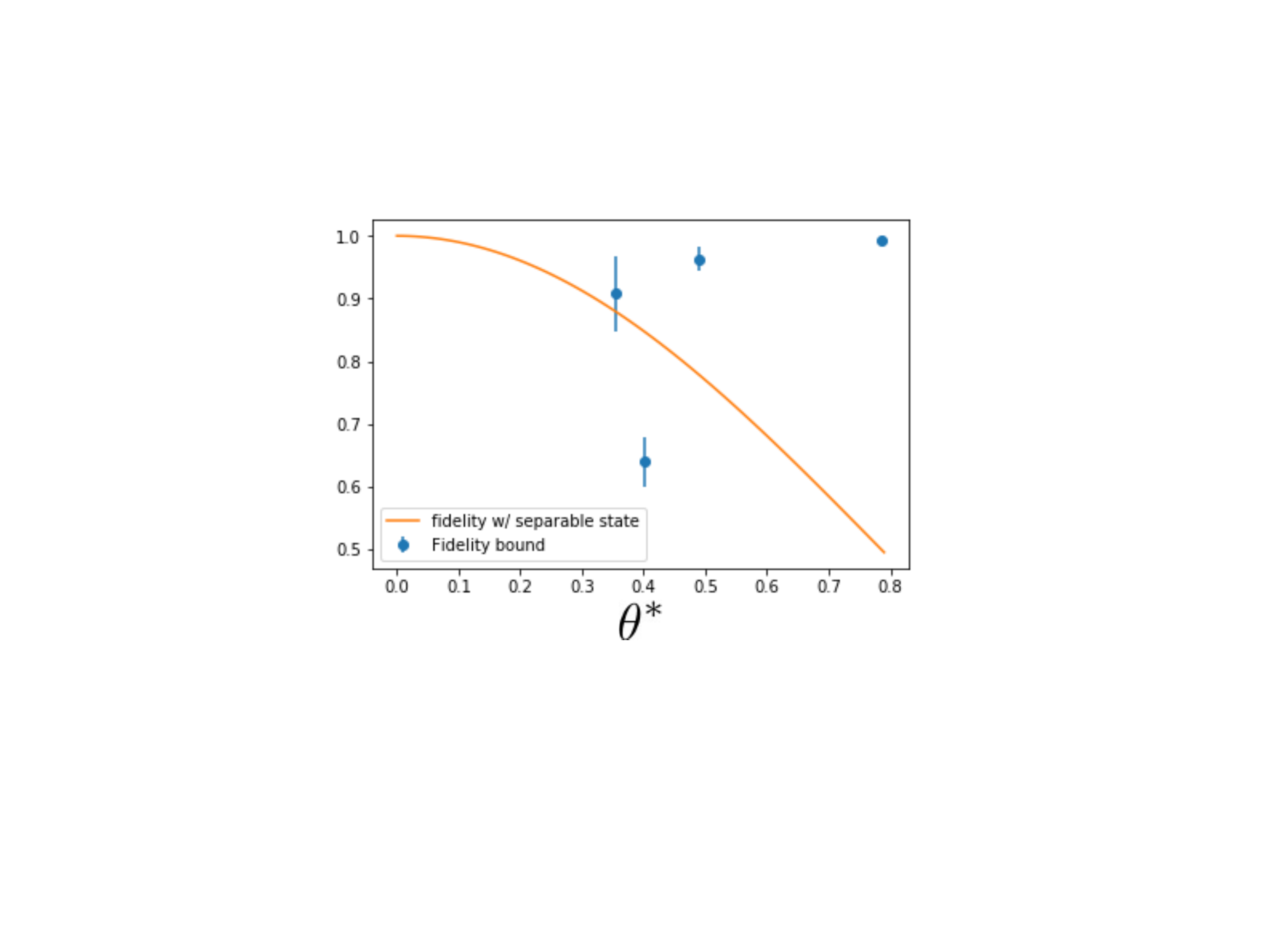}
\caption{Fidelity bounds with respect to target PES. The solid orange line, added as a reference, shows the fidelity of a partially entangled state and a fully separable ones, \ie $F( \ket{00}\bra{00},\ket{\psi(\theta)}\bra{\psi(\theta)})=\cos^2(\theta)$. In the horizontal axis the angle $\theta$ that is self-tested by the fidelity bound. Notice that for the least entangled state the method self-tests a state which is below the separable bound and renders an angle which is very far {from} what obtained through tomography (see Table \ref{t: results}). The error bars are obtained with Gaussian error propagation and considering the Poisson statistics of the recorded coincidence counts.}
\label{fbound}
\end{figure}

Our approach to self-testing uses the bounds to the fidelity with the target PES in Eq. (\ref{fidelitybound}). According to Eq. (\ref{fidelitybound}), it is determined only by the expectation values involved in the violation of the tilted Bell inequality.In our experiment{,} those probabilities were computed for those settings chosen from the estimation of the angle $\theta$. However, when making the DI estimation of the state, all these settings are just labels and the value of $\theta$ estimated through tomography loses its meaning. Again, this is only because the estimation should only be based on the observed correlations. To have a DI estimation of the reference state in the experiment, we optimise the RHS of Eq. (\ref{fidelitybound}) as a function of $\alpha$. Note that the values of all these tilted inequalities can be computed from the observed correlations. The value of $\alpha=\alpha^*$ solving this maximisation provides the searched DI estimation of the angle $\theta^*$, defining the certified closest entangled pure state. The value obtained in the optimisation gives the DI certified fidelity with this state. Table \ref{t: results} and Fig. \ref{fbound} summarise these results. Note the the DI estimation of the fidelities works reasonably well for large values of entanglement, as they are all above 0.9. Moreover, for all these three cases, the DI estimated entanglement, given by $\theta^*$ is reasonably close to the value obtained through tomography. However, for small values of entanglement, the values obtained for the fidelity are not informative despite the high visibilities in the experiment.

\begin{table}[h]
\begin{tabular}{ |c | c | c |c|c|c|}
\hline

   Concurrence & 0.1926 & 0.3746 & 0.5825 & 0.8349 & 0.9858 \\ \hline
   Purity & 0.9849 & 0.9887 & 0.9907 & 0.9846 &  0.9891\\ \hline
   $\alpha$ & 1.914 & 1.741 & 1.373 & 0.949 & 0.0017 \\ \hline
   $B_{\alpha}\pm (0.01)$  & 3.88 & 3.72 & 3.41 & 3.11 & 2.81\\ \hline
   $\frac{B_{\alpha}-L}{Q-L}$ &  -32.84 & -2.14 & 0.78 & 0.91 & 0.98 \\ \hline
   $\theta$  & 0.10539 & 0.19002 & 0.32140 & 0.45946 & 0.7847 \\ \hline
   $\theta^*$ & NA &  0.40059 & 0.35369 & 0.48907 & 0.78536 \\ \hline
      $\epsilon$ & 0.0101 & 0.0143 & 0.0110 & 0.0111 & 0.0126\\
  \hline
 \end{tabular}
  \caption{Summary of experimental results. The concurrence and purity of the states were obtained through quantum state tomography. The coefficient $\alpha$ was chosen as the one providing the best violation of the tilted Bell inequality for the state obtained. $B_{\alpha}$ refers to the observed Bell value, and $\frac{B_{\alpha}-L}{Q-L}$ indicates the relative violation observed ($Q$ and $L$ are the maximum quantum value and the local bound respectively). Notice that the two points with lower entanglement do not violate the inequality. Yet, they are nonlocal and certify randomness through the SDP optimisation. $\theta$ is the angle directly estimated by quantum state tomography, while $\theta^*$ is the angle estimated by self-testing. $\epsilon$ is the difference between the theoretical maximum value that can be obtained and the experimental value observed.}
\label{t: results}
\end{table}

Similar to what happens for randomness certification, the function $B_\alpha$ in Eq. \ref{fidelitybound} increases when decreasing the entanglement of the state, which leads to the increase of the error bars in this regime.

\section{Conclusion}

The theoretical results on self-testing and randomness certification obtained so far indicate that partially entangled states are as good as maximally entangled ones for these tasks in the noiseless scenario. In the present article, we reported on an experimental implementation of these protocols in a photonic experiment. Despite the very low noise levels, the protocols are very sensitive to noise when using weakly entangled states, confirming the intuition that the more the entanglement the better the protocol performance. While of course protocols using weakly entangled states are fragile because they can only produce correlations close to the set of local correlations, it is an interesting open question to understand whether it is possible to construct protocols offering a much stronger performance for a given entangled state. This is a question that deserves further investigation.

\section*{ACKNOWLEDGMENTS}
This work was supported by the ERC CoG QITBOX, Spanish MINECO (QIBEQI FIS2016-80773-P, a Ram\'on y Cajal fellowship, and Severo Ochoa SEV-2015-0522), the AXA Chair in Quantum Information Science, Generalitat de Catalunya (SGR1381 and CERCA Programme), Fundaci\'o Privada Cellex, FONDECYT (1160400 and 11150325), the Millennium
Institute for Research in Optics (MIRO), and PAI-Conicyt (79160083). S.G. acknowledges CONICYT.

\bibliography{Master_Bibtex}

\end{document}